\documentclass[a4paper]{article}

\usepackage{INTERSPEECH2022}

\usepackage{amssymb,amsmath,graphicx}
\usepackage{xspace}
\usepackage{srcltx}
\usepackage{caption}
\usepackage{subcaption}
\usepackage[export]{adjustbox}
\usepackage{multirow,makecell,hhline}
\usepackage{cite}
\usepackage{soul}

\usepackage{url}
\usepackage{hyperref}
\hypersetup{colorlinks=True, urlcolor=black}

\usepackage{color}
\usepackage[dvipsnames]{xcolor}

\newcommand{\CMT}[1]{{}}

\newcommand{\tbh}[1]{\textbf{#1}}

\title{A Language Agnostic Multilingual Streaming On-Device ASR System}
\name{Bo Li, Tara N. Sainath, Ruoming Pang\sthanks{\; Work done in Google.}, Shuo-yiin Chang, Qiumin Xu, Trevor Strohman,}
\nameplus{Vince Chen, Qiao Liang, Heguang Liu, Yanzhang He, Parisa Haghani, Sameer Bidichandani}
\address{Google LLC, USA}
\email{\{boboli,tsainath\}@google.com}

\begin{document}

\maketitle

\begin{abstract}
On-device end-to-end (E2E) models have shown improvements over a conventional model on English Voice Search tasks in both quality and latency.
E2E models have also shown promising results for multilingual automatic speech recognition (ASR). 
In this paper, we extend our previous capacity solution to streaming applications and 
present a streaming multilingual E2E ASR system that runs fully on device with comparable quality and latency to individual monolingual models. To achieve that, we propose an Encoder Endpointer model and an End-of-Utterance (EOU) Joint Layer for a better quality and latency trade-off. Our system is built in a language agnostic manner allowing it to natively support intersentential code switching in real time. To address the feasibility concerns on large models, we conducted on-device profiling and replaced the time consuming LSTM decoder with the recently developed Embedding decoder. With these changes, we managed to run such a system on a mobile device in less than real time.

\end{abstract}

\noindent\textbf{Index Terms}: on-device, multilingual, code switching
\vspace{-0.02in}
\section{Introduction}
\label{sec:intro}
\vspace{-0.05in}

60\% of the world's population speaks more than one language \cite{multilingual}. Most commercial speech recognition systems, however, are designed and optimized in a monolingual setup. 
A common solution is to use multiple recognizers, one for each language, and output the hypothesis with the highest confidence score \cite{gonzalez2014real}. As each recognizer runs independently, the serving cost increases linearly with the number of languages needed, which is impractical for on-device applications. It is also challenging for intrasentential code-switching speech \cite{bullock2009cambridge}, where recognizers of different languages are needed to decode different parts of a single utterance. Alternatively, a language ID (LID) system can be used to detect and split the speech into monolingual segments before recognition \cite{weiner2012integration}, which incurs extra latency. Developing language agnostic multilingual ASR systems is thus critical to enable a natural human-computer speech interaction for the large multilingual user community.

The fast development of end-to-end (E2E) models \cite{Graves2013,Chan15,Chorowski2015AttentionBasedMF,Kim2017,Battenberg2017,JinyuLi2019,Ryan19,Zeyer2020,li2020comparison,graves2012sequence} 
has largely improved automatic speech recognition (ASR) performance, especially for on-device applications. They run directly on users' devices which saves network latency. Users' audio data never leaves the device, which provides better privacy. There have been many successful efforts in building monolingual on-device streaming ASR systems with better quality and latency \cite{QianZhang2020,sainath20streaming,Le2021,bo21system,Shi2021,XieChen2021,sainath2021efficient}.
Multilingual systems, however, have been mainly focusing on non-streaming applications \cite{li2018multi,pratap2020massively,hou2020large,adams2019massively,zhou2018multilingual}. In \cite{kannan2019large}, a streaming Recurrent Neural Network Transducer (RNN-T) \cite{graves2012sequence} system was developed for Indian languages. To address the cross language variations, language dependent adapters was used, which requires knowing the target language information beforehand. \cite{zhou2021configurable} developed a configurable multilingual model that is trained once and can be configured as different language agnostic recognizers. Few works have touched on the latency aspect of such models. In this paper, we extend different work on improving latency \cite{bo21system} to multilingual ASR to fill in this gap. We further propose an Encoder Endpointer model and an End-of-Utterance (EOU) Joint Layer for a better quality and latency trade-off. Besides, we train our multilingual E2E ASR system by pooling data from all the languages without explicitly relying on language information. We find that given a sufficient model capacity, such multilingual systems can outperform monolingual models in both quality and latency. Moreover, these language agnostic multilingual models recognize code-switching speech with lower latency, different from existing work where LID is normally used \cite{lyu2008language,zeng2018end,li2019towards,dalmia2021transformer}. 

\begin{figure}[!t]
\centering
\hspace{-0.34in}
\includegraphics[width=1.1\linewidth]{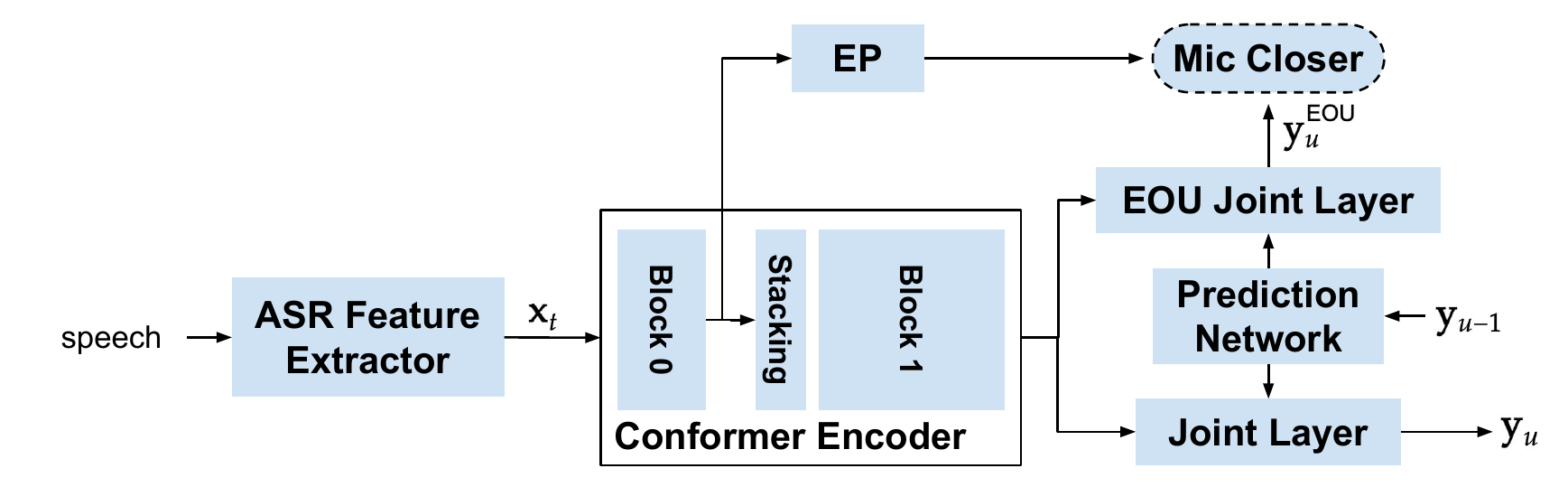}
\caption{{The proposed E2E multilingual streaming ASR system (EP: Endpointer, EOU: End-of-Utterance).}}
\label{fig:system_mmasr}
\vspace{-0.25in}
\end{figure}

\vspace{-0.02in}
\section{System}
\label{sec:system}
\vspace{-0.05in}

In this section, we present the architecture of the proposed multilingual streaming ASR system (Figure \ref{fig:system_mmasr}).

\subsection{E2E ASR Model}

A Conformer \cite{gulati2020conformer} based RNN-T \cite{graves2012sequence} model has recently been shown to outperform a strong hybrid ASR model on English \cite{bo21system,sainath2021efficient}. It consists of an encoder, a prediction network and a joint layer. The encoder consists of multiple Conformer layers \cite{gulati2020conformer} that are arranged into two blocks with a stacking layer in between (Figure~\ref{fig:system_mmasr}). The stacking layer operates in the time dimension by concatenating two adjacent frames to achieve a 2x time reduction. The prediction network is a stack of LSTM layers that summarizes history predictions into a hidden representation. The feed-forward joint layer then combines the encoder and the prediction network outputs to predict a word-piece (WPM) token given the speech inputs. E2E models are often optimized without alignment information and tend to delay token predictions, which can be addressed by FastEmit \cite{yu2021fastemit}. 

In this work, we adopt the same Conformer RNN-T model as \cite{bo21system} for the multilingual system. A truly multilingual system should not only recognize speech from different languages, but also deal well with language/code switching with lower latency. To achieve that, we drop the input language ID \cite{li2018multi,li2021scaling} and use a global input mean and standard deviation normalization and a 16K output WPM obtained from the training data pooled across all the languages.
The large cross-language variation poses a challenge in multilingual training. With the same model capacity, multilingual models tend to perform worse than monolingual models due to the reduced per-language capacity. With larger models, full-context multilingual models can achieve similar qualities as monolingual ones \cite{li2021scaling}. In this study, we further verify this on streaming models.

\subsection{Encoder Endpointer Model}

A streaming ASR system usually uses an {\it Endpointer (EP)} model for better latency (Figure~\ref{fig:system_mmasr}). In Voice Search, an {\it End-of-Utterance (EOU)} EP is normally used. It classifies each input speech frame to be either speech, initial silence, intermediate silence or final silence \cite{chang2017endpoint,chang2019unified}. {\it Mic Closer} then uses this decision to trigger the microphone closing event, which will then end the current speech recognition session and move to the following execution stage. This is referred to as acoustic endpointing \cite{shannon2017improved,chang2020low} and is critical for a low-latency speech experience. 

EP and ASR models are usually developed and maintained independently. Scaling to more languages requires training and maintaining both models. A truly multilingual system requires both the recognizer and the EP model to be multilingual.
To address this, we propose an {\it Encoder Endpointer} model that branches an EP model from the ASR encoder's lower layers. Specifically, as depicted in Figure~\ref{fig:system_mmasr}, the EP model sits on top of the ASR encoder's block 0 before the stacking layer. In this way, EP predictions still have the same frame rate as the input frames while sharing the computation of block 0 with the ASR encoder. This synchronizes the model update for ASR and EP.

\subsection{EOU Joint Layer}

For better latency, RNN-T is normally extended to RNN-T EP \cite{li2020towards} to explicitly emit an End-of-Utterance (EOU) token. This EOU token utilizes ASR's prediction history, brings extra linguistic context information to acoustic EP and has been shown to yield better microphone closing decisions. This is referred to as decoder endpointing \cite{shannon2017improved,chang2020low}. 

Since the EOU token only appears at the sentence end during training, E2E model learns to stop emitting tokens after it. This can cause early cutoffs when inaccurate EOU predictions are made. Besides, we also find the latency benefit of the EOU prediction comes with slight quality degradations \cite{chang2019joint, li2020towards}. 
In this paper, we propose a simple change to the RNN-T EP model by using an additional joint layer, namely the EOU Joint Layer. As illustrated in Figure~\ref{fig:system_mmasr}, we keep the original joint layer unchanged and add an extra joint layer for the prediction of word-piece tokens together with EOU. To ensure the same 
recognition quality, we conduct the training in two stages. Firstly, we optimize all the encoder, prediction network and the original joint layer for recognition quality; then we freeze all those parameters and initialize a new EOU Joint Layer from the original joint layer and fine-tune only this EOU Joint Layer. During inference, we rely on the original joint layer for beam search decoding and only use the EOU Joint Layer to generate the EOU probability.

\section{Experimental Setup}
\label{sec:exp}

\subsection{Data}

\begin{table}[t]
\caption{Per language training data statistics. Utterance counts are in millions (M) and duration is in thousand (K) hours.}
\centering
\vspace{-0.1in}
\begin{tabular}{llrr}
\toprule
\tbh{Locale} & \tbh{Language} & \tbh{Counts}(M) & \tbh{Hours}(K)\\
\midrule
\midrule
en-US & English (USA) & 34.0 & 52.6 \\
zh-TW & Chinese (Taiwan) & 16.6 & 22.0 \\
fr-FR & French & 16.5 & 23.7 \\
de-DE & German &  15.3 & 23.4 \\
ja-JP & Japanese & 14.9 & 20.2 \\
es-US & Spanish (USA) & 14.2 & 23.8 \\
es-ES & Spanish (Spain) & 12.9 & 20.1 \\
it-IT & Italian & 11.8 & 19.8 \\
en-GB & English (UK) & 6.0 & 8.6 \\
\midrule 
\midrule
Total & & 142.3 & 214.2 \\
\bottomrule
\end{tabular}
\label{tbl:data}
\vspace{-0.2in}
\end{table}

We use a dataset consisting of speech from 9 Voice Search locales with a total number of 142.3M utterances. The total duration is around 214.2K hours. The data is anonymized and human transcribed. The per language data distribution is listed in Table \ref{tbl:data}. The number of utterances for each language ranges from 6.0M to 34.0M, roughly corresponding to 8.6K to 52.6K hours of speech data. 
The test set for each language contains around 3.0K-15.4K utterances less than 5.5s in length (Table~\ref{tbl:results_all}). They are also sampled from Voice Search traffic with no overlap from the training set. Similarly, they are anonymized and hand-transcribed for evaluation purpose.

\begin{table*}[!ht]
\caption{WER (\%) comparisons of different systems without Endpointer (EP).}
\centering
\vspace{-0.1in}
\resizebox{0.95\textwidth}{!}{
\begin{tabular}{lrcccccccccc}
\toprule 
\multirow{2}{*}{\tbh{System}} & \multirow{2}{*}{\tbh{Size \scriptsize{(M)}}} & \multicolumn{9}{c}{\tbh{Per Language WER \scriptsize{(\%)}}} & \tbh{Avg.} \\ 
\cmidrule(lr){3-11}
~ & ~ & \tbh{en-US} & \tbh{zh-TW} & \tbh{fr-FR} & \tbh{de-DE} & \tbh{ja-JP} & \tbh{es-US} & \tbh{es-ES} & \tbh{it-IT} & \tbh{en-GB} & \tbh{WER \scriptsize{{(\%)}}} \\
\midrule
\midrule
Utterance Counts \scriptsize{(K)} & - & 5.0 & 15.4 & 3.4 & 12.0 & 14.1 & 3.0 & 12.5 & 14.2 & 7.9 & - \\
\midrule
Monolingual \scriptsize{Conformer RNN-T} & 140 * 9 & 6.3 & 4.7 & \bf{12.5} & 12.9 & 10.5 & 7.8 & 7.3 & 7.2 & 7.2 & 8.49 \\
\midrule
S1 \scriptsize{with input LID} & 140 & 6.1 & 5.2 & 12.6 & 13.3 & 11.9 & 7.0 & 6.9 & 7.7 & 6.1 & 8.53 \\
S2 \scriptsize{truly multilingual} & 140 & 6.3 & 5.4 & 12.8 & 13.6 & 12.0 & 7.2 & 7.1 & 8.1 & 6.3 & 8.76 \\
\midrule
S3 \scriptsize{S2 with FastEmit} & 140 & 7.0 & 5.6 & 12.8 & 14.0 & 12.5 & 7.4 & 7.3 & 8.5 & 6.5 & 9.07\\
S4 \scriptsize{S3 with EOU} & 140 & 7.5 & 6.1 & 13.2 & 14.4 & 13.0 & 7.9 & 7.6 & 9.0 & 7.0 & 9.52 \\
\midrule
S5 \scriptsize{deeper and wider S3} & 500 & 6.1 & 4.9 & 12.7 & 13.2 & 11.1 & \bf{6.9} & 6.8 & 7.2 & 5.9 & 8.31 \\
S6 \scriptsize{even wider S5} & 1000 & \bf{5.6} & \bf{4.8} & 12.6 & \bf{12.6} & \bf{10.4} & 7.0 & \bf{6.5} & \bf{6.6} & \bf{5.6} & \bf{7.97} \\
\midrule
S7 \scriptsize{S6 with Embedding decoder} & 920 & 6.1 & 5.0 & \bf{12.5} & 12.8 & 11.3 & 7.0 & 6.7 & 7.1 & 5.9 & 8.27 \\
\bottomrule
\end{tabular}}
\label{tbl:results_all}
\end{table*}

\subsection{Model Architecture}
\label{sec:model}

We use 80-dimensional(-dim) log Mel filterbank features that are computed on 32ms windows with a 10ms hop. Features from 3 contiguous frames are stacked and sub-sampled to form a 240-dim input representation with 30ms frame rate. SpecAugment \cite{park2020specaugment} is used to improve models' robustness against noise. Specifically, two frequency masks with a maximum length of 27 and two time masks with a maximum length of 50 are used. 

Following \cite{sainath2021efficient}, we use 512-dim Conformer layers in the ASR encoder. Causal convolution and left-context attention layers are used for the Conformer layer to strictly restrict the model to use no future inputs. 8-head attention is used in the self-attention layer and the Convolution kernel size used is 15. The encoder consists of 12 Conformer layers, separated into two blocks by a stacking layer. Block 0 consists of an input projection layer and 3 Conformer layers. The time stacking layer concatenates two adjacent frame outputs from block 0 with no overlap to form a 60ms frame. After that, block 1 first uses a 1024-dim Conformer layer, then a projection layer to reduce the model dim back to 512 and lastly 8 Conformer layers followed by a layer normalization. The RNN-T decoder consists of a prediction network with 2 LSTM layers with 2,048 units projected down to 640 output units, and a joint network with a single feed-forward layer of 640 units. All the E2E ASR models are trained to predict 16,384 word pieces generated from the transcripts pooled across all the languages. The final model has 110M parameters in the encoder and 33M parameters in the decoder. It is optimized by minimizing the RNN-T loss. FastEmit \cite{yu2021fastemit} regularization with a weight of 5e-3 is used to reduce prediction latency.

For the encoder endpointer model, we firstly project down the ASR encoder block 0's output to 128-dim and then use a single layer of 128-dim Conformer layer. The output is further project down to 4-dim and layer normalized before the final softmax layer. The model has 449K additional parameters specific for endpointing. Per-frame cross entropy loss is used to train these endpointer specific parameters.

All the models are trained in Tensorflow using the Lingvo toolkit \cite{shen2019lingvo} on Google's Tensor Processing Units (TPU) V3 \cite{tpu} with a global batch size of 4,096 utterances. Models are trained with 512 TPU cores and optimized using synchronized stochastic gradient descent. We use the Adam optimizer \cite{shazeer2018adafactor} with parameters $\beta_1$=0.9 and $\beta_2$=0.999. A transformer learning rate schedule with peak learning rate 1.8e-3 and 32K warm-up steps is used. Exponential moving average \cite{Vaswani17} has been used to stabilize the model weight updates. 
\section{Results}
\label{sec:results}

\subsection{A Truly Multilingual ASR System}

\begin{figure}[ht]
\hspace{-0.1in}
    \includegraphics[scale=0.38]{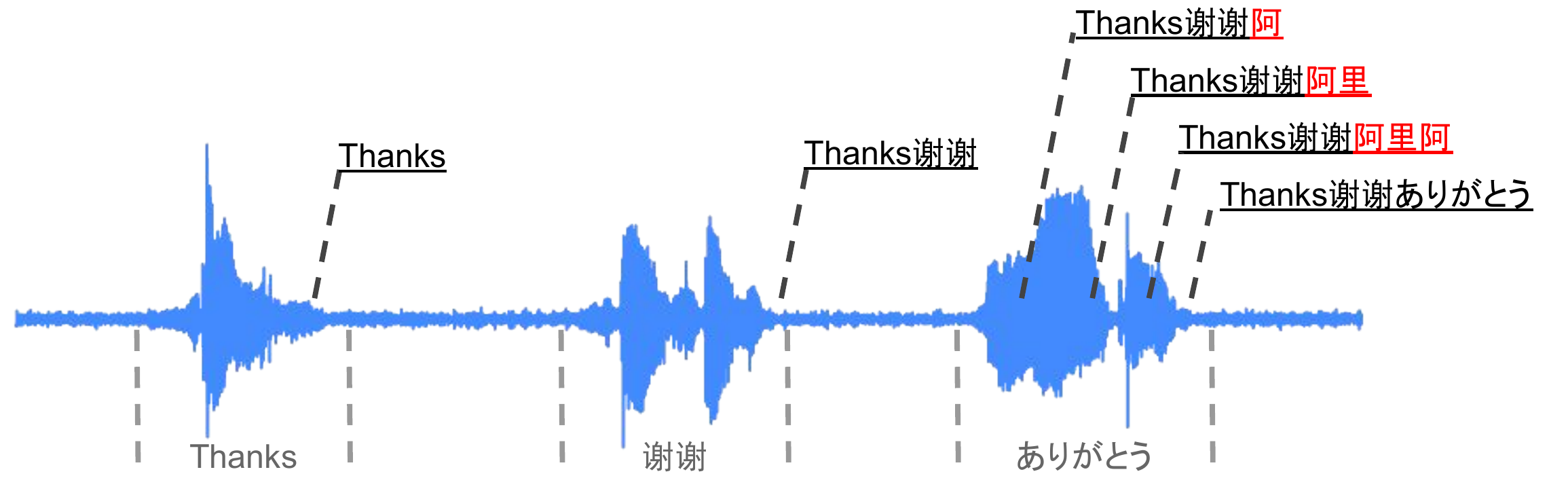}
    \vspace{-0.1in}
    \caption{A code-switching example: a user speaking ``thanks'' in English, Chinese and Japanese. Texts below the waveform are the ground truth and those above are the top partial hypotheses at corresponding time (not all top partial hypotheses are plotted).}
    \vspace{-0.1in}
    \label{fig:demo}
\end{figure}

To build a truly multilingual system (S2 in Table~\ref{tbl:results_all}), we pool all the data from different languages together with its natural distribution. Different from previous work \cite{li2018multi, li2021scaling}, we do not feed any language information into the model. We use the architecture described in Section \ref{sec:model}, specifically a 12-layer Conformer encoder and 2-layer LSTM decoder streaming RNN-T model. 
We compare with two baselines: 1) monolingual models (Monolingual in Table~\ref{tbl:results_all}) and 2) a multilingual model with a 16-dim 1-hot LID vector concatenated to the filterbank features (S1 in Table~\ref{tbl:results_all}). All the three systems use the same model architecture as described in Section~\ref{sec:model}, except that the monolingual models use language dependent WPM outputs. From the results in Table~\ref{tbl:results_all}, multilingual models degrade on languages with the most data but outperforms monolingual models on es-US, es-ES and en-GB which have slightly less data. Oracle language information (S1) brings quality gains on monolingual test sets, but limits the capability in recognizing speech with code-switching phenomenon. For example, to recognize an utterance with different languages (Figure~\ref{fig:demo}), S1 requires language ID information before recognition. However, with S2 we can natively switch language during the recognition. We hence take the multilingual model without language ID (S2) for the following studies.

Due to the lack of proper code-switching test sets, we conducted some preliminary user studies to understand S2's capability in handling such speech. It performs surprisingly well despite the fact that we do not have code-mixing data during training. In one example as shown in Figure~\ref{fig:demo}, S2 mistakenly outputs Chinese characters after the user switches from Chinese to Japanese (red colored texts in Figure~\ref{fig:demo}), but it manages to correct itself after seeing more Japanese speech and outputs the correct transcript at the end. Future work will investigate intrasentential code switching in great details. 

\begin{table}[t]
\caption{Final silence accuracy (\%) of different EOU Endpointer models.}
\centering
\vspace{-0.1in}
\resizebox{0.45\textwidth}{!}{
\begin{tabular}{llrr}
\toprule
\tbh{Model} & \tbh{Architecture} & \tbh{Size(K)} & \tbh{Accuracy(\%)}\\
\midrule
\midrule
EP0 & Standalone (3-layer LSTM) & 462.5 & 91.6\\
\midrule
EP1 & Encoder Endpointer & 2.0 & 88.3\\
EP2 & EP1 + 3-layer LSTM & 462.5 & 91.7 \\
EP3 & EP1 + 1-layer Conformer & 448.7 & 93.8\\
\bottomrule
\end{tabular}}
\label{tbl:endpointer}
\vspace{-0.1in}
\end{table}

\begin{table*}[!ht]
\caption{Quality and Latency (EP50: 50-percentile and EP90: 90-percential) comparisons between the proposed multilingual model (S3) and classical hybrid monolingual models.}
\centering
\vspace{-0.1in}
\resizebox{0.95\textwidth}{!}{
\begin{tabular}{llcccccccccc}
\toprule 
\multirow{2}{*}{\tbh{Metric}} & \multirow{2}{*}{\tbh{System}} & \multicolumn{9}{c}{\tbh{Per Language}} & \multirow{2}{*}{\tbh{Avg.}} \\
\cmidrule(lr){3-11}
~ & ~ & \tbh{en-US} & \tbh{zh-TW} & \tbh{fr-FR} & \tbh{de-DE} & \tbh{ja-JP} & \tbh{es-US} & \tbh{es-ES} & \tbh{it-IT} & \tbh{en-GB} & ~ \\
\midrule
\midrule
\multirow{2}{*}{\tbh{WER \scriptsize{(\%)}}} & Monolingual \scriptsize{classic} & \bf{7.5} & \bf{4.9} & \bf{12.9} & \bf{13.7} & \bf{10.8} & 9.3 & \bf{8.1} & \bf{8.2} & \bf{6.4} & \bf{~9.09} \\
~ & S3 & 11.0 & 6.6 & 15.0 & 16.4 & 15.1 & \bf{8.7} & 9.3 & 11.3 & 9.8 & 11.47 \\
\midrule
\multirow{2}{*}{\tbh{EP50 \scriptsize{(ms)}}} & Monolingual \scriptsize{classic} & 430 & 370 & 380 & 300 & 510 & 440 & \bf{440} & 540 & 510 & 436 \\
~ & S3 & \bf{330} & \bf{450} & \bf{470} & \bf{200} & \bf{420} & \bf{320} & 450 & \bf{530} & \bf{350} & \bf{391} \\
\midrule
\multirow{2}{*}{\tbh{EP90 \scriptsize{(ms)}}} & Monolingual \scriptsize{classic} & 880 & 780 & 720 & 740 & 890 & 840 & 820 & 990 & 960 & 847 \\
~ & S3 & \bf{670} & \bf{630} & \bf{660} & \bf{570} & \bf{780} & \bf{690} & \bf{670} & \bf{800} & \bf{730} & \bf{689}\\
\bottomrule
\end{tabular}}
\label{tbl:final}
\end{table*}

\subsection{Reducing Latency}

Besides quality, latency is another critical metric for streaming applications. We compare different endpointer models for the streaming multilingual ASR system. We report only the frame accuracy for the final silence class which is used in the recognition pipeline for microphone closing decisions. We first train a conventional standalone 3-layer 128-dim LSTM endpointer (EP0 in Table~\ref{tbl:endpointer}) \cite{chang2019unified} as our baseline. For the proposed encoder endpointer, we explored a few different architectures including: a single projection layer (EP1), EP1 with additional 3-layer 128-dim LSTM (EP2), and EP1 with a comparable size but using a single Conformer layer (EP3). The per frame final silence prediction accuracy of these models together with the number of additional parameters brought by these EP models are listed in Table~\ref{tbl:endpointer}. EP1 has the lowest accuracy suggesting the need for endpointer specific model parameters. Simply projecting down the ASR encoder's block 0 output for endpointing is not sufficient. Another potential solution is to jointly train the whole system with both ASR and endpointer targets. This way the encoder block 0 is aware of the endpointer task in training, which will be explored in future work. Next, with the same number of additional parameters (EP0 vs. EP2), encoder endpointer yields better accuracy, benefiting from the pretrained ASR encoder. Lastly, after switching from LSTM to Conformer, we obtained the best 93.8\% final silence prediction accuracy.

E2E models trained without explicit alignment constraints tend to delay output predictions, which increases the latency for streaming applications. To address that, we adopted FastEmit \cite{yu2021fastemit} in our multilingual training. Similar to existing findings \cite{bo21system}, FastEmit hurts WER (comparing S3 to S2 in Table~\ref{tbl:results_all}).
Besides reducing RNN-T's inherent token prediction latency, we also utilize decoder endpointer to reduce the microphone closing latency. We append an EOU token to the end of each training WPM sequence. On top of FastEmit, we found no additional late penalty \cite{li2020towards} is needed for the EOU prediction. From the results (S4 in Table~\ref{tbl:results_all}), we can see adding EOU token further degrades the recognition quality. To address this problem, we move the the EOU prediction into a separate EOU Joint Layer and keep the existing RNN-T model unchanged. In this way, we can maintain the same quality as S3 while still benefiting from the EOU prediction for a fast decoder endpointing.

\subsection{Comparison with Monolingual Models}

With all the above components, we conducted an end-to-end test with both endpointer and ASR models. For monolingual RNN-T models (Monolingual in Table~\ref{tbl:results_all}), we need to conduct language specific parameters tuning to find the best quality and latency trade-offs for a fair comparison, which is a time consuming task. Instead, we compared with the current best production models for Voice Search, which are extensively tuned hybrid monolingual ASR systems (Monolingual in Table~\ref{tbl:final}) consisting of acoustic models, lexicons and language models \cite{sainath2021efficient}. The final quality and latency metrics are depicted in Table~\ref{tbl:final}. EP is crucial to streaming application latency but hurts WER (comparing the WER for S3 in Table~\ref{tbl:final} to Table~\ref{tbl:results_all}). Our single 140M-param multilingual model (S3) lags behind monolingual models, except that on es-US, S3 achieves better quality. Moreover, the latency of the proposed multilingual system S3 is much better, especially in terms of the 90th percentile latency (EP90 in Table~\ref{tbl:final}).

\subsection{Further Quality Improvements}

To address the quality gap between the proposed multilingual model and the monolingual models, we hypothesize the main problem is the limited model capacity \cite{li2021scaling}. The same number of model parameters as each monolingual model now needs to capture more variations across languages. Similar to \cite{li2021scaling}, we increase S2's encoder width from 512 to 640 and depth from 12 to 35, and decoder width from 2048 to 3072 and depth from 2 to 6, leading to a 500M-param model S5. By further increasing the encoder width to 1024, it brings us a 1B-param model S6. From Table~\ref{tbl:results_all}, with 1B-param model, we can achieve better or comparable performance to each monolingual models. In terms of endpointer latency, namely EP50 and EP90, the larger models perform similarly to S3 in Table~\ref{tbl:final}. 

\subsection{On-Device Benchmark}

To test the feasibility of deploying these large models to mobile devices, we conducted some preliminary benchmark studies on Google Pixel 6 phones with Google Tensor \cite{googletensor} on 100 utterances. We report the on-device real-time factor and memory usage to show the computation efficiency and resources needed to serve such models (Table \ref{tbl:rt}). For the 1B-param model (S6), it is impractical for streaming applications. However, after detailed analysis, we found the LSTM decoder takes up most of the computation. Recent work \cite{prabhavalkar2021less,botros2021tied} has found that stateless decoders are sufficient for E2E models. We hence swapped the LSTM decoder of S6 with a 33M-param Embedding decoder (S7 in Table \ref{tbl:results_all} and Table~\ref{tbl:rt}). It increases the WER but is still better than most monolingual models (Table~\ref{tbl:results_all}). More importantly, it largely reduces the real-time factor and enables the deployment of such large models to mobile devices.

\begin{table}[h]
\caption{On-device benchmark for different sized models (RT50: Real-time factor, 50th percentile, RT90: Real-time factor, 90th percentile and MEM: peak memory usage).}
\centering
\vspace{-0.1in}
\resizebox{0.45\textwidth}{!}{
\begin{tabular}{cccccc}
\toprule
~ & \tbh{140M(S2)} & \tbh{500M(S5)} & \tbh{1B(S6)} & \tbh{920M(S7)} \\
\midrule
\midrule
RT50 & 0.43 & 1.76 & 1.96 & 0.61 \\
RT90 & 0.80 & 2.70 & 3.05 & 0.70 \\
\midrule
MEM(G) & 0.38 & 1.08 & 1.63 & 1.03 \\
\bottomrule
\end{tabular}}
\label{tbl:rt}
\vspace{-0.2in}
\end{table}
\section{Conclusions}
\label{sec:concl}

In this paper, we presented a language agnostic multilingual streaming on-device end-to-end system with comparable quality and latency metrics to monolingual models. On the quality side, we found a larger model capacity is needed to obtain better qualities than monolingual models. Our on-device benchmark study showed that it is feasible to run large models in less than real time on a modern mobile device. On the latency side, we developed an Encoder Endpointer multitask model to unify the training of the multilingual endpointer and the speech recognizer. We further proposed to use a separate End-of-Utterance (EOU) Joint Layer which reduces the endpointing latency with EOU prediction while maintaining the recognition quality. With the fast advancement in both software and hardware, deploying high-quality and low-latency multilingual on-device speech experience would become practical in the near future.

\bibliographystyle{IEEEtran}
\bibliography{main}

\end{document}